# Linear Exp-6 Isotherm for Compressed Molten Cesium over the Whole Liquid Range Including Metal-nonmetal Transition and $T_c$

M.H. Ghatee[*], and H. Shams-Abadi

*(Department of Chemistry, Shiraz University, Shiraz 71454, Iran)*

E-mail: ghatee@sun01.susc.ac.ir

## ABSTRACT

*The Linear exp-6 isotherm is presented as an approach to the thermodynamic properties of liquid alkali metals over the whole liquid range including metal-nonmetal transition. The exp-6 pair interaction potential is applied to approach the underlying interplay between the characteristics soft repulsive interaction in dense, large attractive interaction in expanded liquid alkali metal and the observed thermodynamic properties. PVT of a dense liquid alkali metals obey the linear exp-6 isotherm* $(Z-1)V^2 = A + B\rho^{-7/3}\exp[\alpha(1 - C\rho^{-1/3}/r_m)]$ *over the whole range of liquid densities, where* $Z$ *is the compression factor,* $\rho = 1/V$ *is the density,* $r_m$ *is the position of potential minimum and* $\alpha$ *is a parameter and C is a constant. The intercept A and the slope B significantly are related to attraction and repulsion, respectively, and both depend on temperature.*

*At the level of theory presented in this study, the thermodynamic effects caused by the polarization of atoms in expanded liquid alkali metals can be accounted for by exp-6 potential function to demonstrate their thermodynamic properties as normal liquids. This includes the metal-nonmetals' transition range in which case the nature of forces is changed in such a way that the thermodynamic properties would be different from those of the high-density region. In particular, the equation of state for molten cesium is analytical over the whole range of liquid densities. The linear exp-6 isotherm is used to estimate the binding energy (0.0733eV) and the position of potential minimum (* $5.336 \dot{A}$ *) of liquid cesium at freezing temperature.*

## 1. Introduction

Alkali metals, both in liquid and in vapor states, are complicated in structure and in molecular (clusters) interaction. In their vapor states, they are spectroscopic samples for investigation of electronic structures and interatomic interaction, and they are examples for ease of ab initio calculations. In liquid state, they are useful for their marked thermodynamic properties such as high heats of vaporization and large liquid ranges, which make them good heat transfer fluids in reactors operating at high temperature and at high-energy rate, as high as nuclear power reactors. Liquid alkali metals are also good conductors of electrical current[1]. The extent of electrical conduction, however, is decreased as the liquid metal is expanded and drops sharply at the characteristic density. Such information on physical and thermodynamic properties of alkali metals is important in technological application of these groups of elements. These wide peculiar applications prompted the investigation of the details of structure and interaction at the molecular level.

Liquid alkali metals have been treated thermodynamically by methods of dense fluids[1]. Due to the overlap of wave function of single valence electrons, the highly effective interatomic interaction results in a soft-referenced potential. In particular, the heavier alkali metals are known by their shallow potential well. On the other hand, liquid alkali metals' atoms (especially at high temperature) are readily polarized so much that the structure of their attractive side of the related potential functions is different from those of normal fluids[2]. The polarization could, however, reject the assumption of radial interatomic force field. As the critical point is approached, due to the gradual localization of electrons and the subsequent metal-nonmetal transition, liquid alkali metals gradually turn to insulators and their thermodynamic behaviors would be different from those at low temperature (i.e., high density).[3] Alkali metals also violate the law of corresponding states, which assumes that the interaction between fluid particles has the same form in liquid, vapor, and gas states. This is largely true for molecular fluids such as Ar and Xe because the type of binding is similar in liquid and vapor. As a result of these diversions of the nature of forces, the potential energy (function) would depend on the electronic state of the bulk of liquid metals and thus the thermodynamic treatment for physical properties requests a more elaborated technique.

In a recent study, we have applied the results of linear regularity isotherm (LIR) to molten alkali metals.[4] (LIR simply states that for dense normal fluids,[5] isotherms of $(Z-1)V^2$ are linear in $\rho^2$, where Z is the compression factor, V is volume, and $\rho = 1/V$). Additionally, we have explored isotherms of molten Cs close to critical temperature and have attributed the deterioration of the linearity, which we have observed in this region, to an alteration in the assumption of nearest neighbor in the basic derivation of LIR.[4,5] Standing in the favor of this unbiased viewpoint, in a cause and effect type article,[6] liquid alkali metals have, after the transition, been analyzed as an ensemble of mixture of distinguishable monomers and dimers. Then, keeping the track of a Lennard-Jones model potential[5] (see results and discussion), the system composition in terms of temperature and density has been followed and it was finally established that a LIR EOS is quadratic in $\rho^2$. They finally argued that, the linear isotherm persists at any temperature unless the densities of liquid alkalis are smaller than their characteristic density limits at which n-mers are being formed and, thus, the quadratic behavior is started.[6] The density limit corresponds to the density
below which a marked decrease in the electric conductivity in liquid alkali metal is occurred, $1.21 \, \text{gr/cm}^3$ for Cs, for instance.[3]

Favoring this viewpoint, we will pursue the problem of the deterioration of the linearity[4] of the linear isotherms that are observed as a result of metal-nonmetal transition in the expanded liquid alkali metals close to critical temperature.[6] We will consider the interatomic interaction of the liquid alkali metals that distinguishes them from normal liquids; namely, a more soft repulsion at short range and a larger attraction at long range (than the normal liquid), due to overlap of valence electrons and polarization of atoms, respectively. We believe that modeling the potential of the system in both ranges in accord with the diversion of forces will allow an accurate estimation of thermodynamic properties and, thus, lead to an exact EOS.

The purposes of this study are, (*i*) to investigate the application of exp-6 potential to model the interactions in dense and expanded liquid alkali metals, (*ii*) to evaluate the range of accuracies of this potential function as a pre-assumed model potential in construction of the linear isotherm, and (*iii*) to infer an equation of state that suitably augments the effects of internal diversion of electronic forces and subsequent metal-nonmetal transition as the critical temperature is approached. Of all alkali metals, experimental data have measured over the whole range of temperature only for liquid Cs to which our study is mainly referred.[7,8]

## 2. Interaction potential

The single valence electrons of alkali metal atoms cause the repulsive interaction to become softer than those of normal liquids. Also, at high temperatures, the (sizable) attractions are due to polarization of alkali metal atoms. For spectroscopic studies (in vapor state), the Morse potential function has been shown to furnish large attractions at long range. The attraction was also modeled in the form of dispersive multipole expansion.[9] The pair potential functions for alkali metals from spectroscopic results basically corresponds to zero density limit and their applications to liquid state at most lead to a moderate accuracy. The asymptotic interaction behaviors in molten alkali metals to some extent coincide with the exp-6 pair potential (eq 1) and we will apply it as a pair potential function to model the thermodynamic properties of these metals.

$$u(r) = \varepsilon\left\{\frac{6}{\alpha-6}\exp\left(\alpha\left(1-\frac{r}{r_m}\right)\right) - \frac{\alpha}{\alpha-6}\left(\frac{r_m}{r}\right)^6\right\} \qquad r > r_{max}, \qquad (1)$$

$$u(r) = \infty \qquad r \leq r_{max}.$$

In eq 1, $r_{max}$ is the short distance where the potential function tends toward large negative values, $\varepsilon$ is the potential well depth with $r_m$ being position minimum. The parameter $\alpha$ adjusts the slope of the repulsive as well as the attractive side of the potential function. The smaller the value of $\alpha$ the exp-6 potential function models a fluid of the softer repulsive interaction and of the larger attractive interaction. This is graphically depicted in Figure 1 and compared with the Lennard-Jones potential function. Recently, exp-6 potential has found wide applicabilities in modeling and simulations of dense fluids.[10]

The slopes of repulsive side of the potential functions of a number of fluids like $O_2, N_2, CH_4, CO,$ and $CO_2$ have been estimated by considering the values of $\alpha$, equal to 13 for all of them. Thus, they obey the law of corresponding states for a number of physical properties.[11] The applications of eq 1 for theoretical and practical purposes have already been established.[11] The attraction potential of the eq 1 can be adjusted so that the polarizable systems can be quantitatively treated.[12]

Polarizablities in the alkali metals are more pronounced at high temperatures. It is such a controversial matter that upon laser excitation alkali metals' vapor is dimerized readily by penetration of one atom inside the valence shell of another highly excited atom so that an ionized (or partially ionized) dimer is produced.[13,14] Notably, Rydberg and Morse potentials suitably fit to IR spectroscopic data because of their capability of a better estimate of the attractions than the simple Lennard-Jones potential. More recently, pair potentials for alkali metals at different molecular electronic states have been presented by Konowalow.[15]

Attempts always have been to reproduce molecular parameters from macroscopic properties and the manipulation of statistical thermodynamic.[16,17] It seems to be much restrictive by using liquid PVT data in assessment of molecular parameter. This is due to the fact that a dense fluid is more structured by the effective repulsion of the neighboring molecules in the force field range of the pair whose interaction potential is being considered.

### 3. Exp-6 Isotherm

Details of interaction potential function and the structure of liquid state are important in construction of an accurate analytical equation of state. Practically, the thermodynamic equation of state can be solved to estimate the pressure of dense fluids from the knowledge of pair potential function and the structure of the fluid. The potential energy U of N molecules confined in volume V can often be fairly well approximated by a pairwise sum of pair potentials

$$U(\mathbf{r}_1,\cdots,\mathbf{r}_N) = \sum_{i>j}^{N} u(\mathbf{r}_i,\mathbf{r}_j) \qquad (2)$$

where the pair potentials $u(\mathbf{r}_i,\mathbf{r}_j)$ are often assumed to depend only upon the distance $\mathbf{r}_{ij} = |(\mathbf{r}_i - \mathbf{r}_j)|$ between the *i*th and *j*th pair of molecules located at positions $\mathbf{r}_i$ and $\mathbf{r}_j$, respectively. This approximation, which simplifies the evaluation of thermodynamic properties from formulation of statistical mechanics, is too crude to describe the thermodynamic behaviors of dense fluid systems. In this study, we assume a complete pairwise additive of the potential, however, in a different sense by assuming that all of the N molecules in pairs interact like molecule 1 interacting with molecule 2 at distance $r_{12}$ with pair potential $u(\mathbf{r}_{12})$ and, thus, we calculate the potential energy as

$$U(\mathbf{r}_1,\cdots,\mathbf{r}_N) = \frac{N}{2}u(\mathbf{r}_{12}) \qquad (3)$$

where the factor $N/2$ is the number of dissimilar pairs of interactions. Often $u(r_{12})$ is represented by $u(r)$, where r is the distance of two nearest neighboring atoms and the pair potential of the type Eq. 1 can be applied. This choice of total potential energy will find reality when experimental liquid PVT data are used to assess the molecular parameters. Thus, in the sense of this modeling the potential well depth $\varepsilon$ of the pair potential would turns out to be the binding energy of pair of molecules 1 and 2 in the ensemble of N-2 other molecules. The most effective interactions are due to the nearest neighbors. Interestingly, the binding energy depends on the choice of pair potential function used to model the fluid system, because a given potential model introduces an effective characteristic interaction range.

The total internal energy E of the fluid is the sum of potential energy U and the kinetic energy K. The kinetic energy represents the thermal energy of the fluid and thus is a function of temperature. Inserting eq 1 into eq 3 and differentiating E with respect to volume $V \propto r^3$ algebraically yields the internal pressure $P_{int}$ in terms of molecular potential parameter $\varepsilon, r_m$, and V :

$$P_{int} = \left(\frac{\partial(U+K)}{\partial V}\right)_{N,T} \approx -B_1 V^{-2/3} \exp\left[\alpha\left(1 - \frac{CV^{1/3}}{r_m}\right)\right] + A_1 V^{-3} \qquad (4)$$

where,

$$B_1 = \frac{CN\varepsilon}{r_m} \frac{\alpha}{(\alpha-6)}, \quad A_1 = \frac{N\varepsilon r_m^6}{C^6} \frac{\alpha}{(\alpha-6)}, \qquad (5)$$

and $C = \left(3\sqrt{3}/4N\right)^{1/3}$. Notice that the solid alkali metals are crystallized in body-centered cubic unit cells, and their molten structures can be well approximated by the same crystalline form. The contribution of kinetic energy K to the total energy vanishes in the manipulations because it is fairly independent of volume. It is noticeable that the internal pressure is approximated by the interactions due to dissimilar pairs of nearest neighbors whose effective range of interaction is determined by the pair potential function eq 1.

The pressure of the system can be calculated by using the well-established exact thermodynamic equation of state:

$$P = T\left(\frac{\partial P}{\partial T}\right)_{N,V} - P_{int}. \qquad (6)$$

Inserting eq 4 into eq 6 gives the pressure in terms of molecular parameters and the molar volume, and accordingly the equation of state reads as

$$(Z-1)V^2 = \frac{1}{\rho^2}\left[\frac{1}{R\rho}\left(\frac{\partial P}{\partial T}\right)_{N,V} - 1\right] + \frac{B_1}{RT} V^{7/3} \exp\left[\alpha\left(1 - \frac{CV^{1/3}}{r_m}\right)\right] - \frac{A_1}{RT}. \qquad (7)$$

where Z is compression factor in molar unit. It can then be rearranged to obtain the equation of state

$$(Z-1)V^2 = A + B\rho^{-7/3} \exp\left[\alpha\left(1 - \frac{C\rho^{-1/3}}{r_m}\right)\right], \qquad (8)$$

where the molar density $\rho = 1/V$, A and B are two temperature-dependent constants of the equation of state:

$$A = \frac{1}{\rho^2}\left[\frac{1}{R\rho}\left(\frac{\partial P}{\partial T}\right)_{N,V} - 1\right] - \frac{A_1}{RT} = A_2 - \frac{A_1}{RT}, \quad B = \frac{B_1}{RT}. \qquad (9)$$

In above derivations N is taken as Avogadro's number to switched to molar units. To derive eq 8, we have closely followed the method of ref 5. But the major difference is the employment of the pre-assumed exp-6 potential model and summing only for the potential energy of dissimilar pairs of nearest neighbors. Our approach treats the potential well depth as the binding energy of the pair of molecules 1 and 2 in ensemble of N-2 identical molecules. Interestingly, the employment of Lennard-Jones potential function as a pre-assumed pair potential model by the procedure just outlined leads to the same results of the ref 5.

## 4. Results and Discussion
**4.1. Exp-6 Isotherms.** The regularity in our treatment predicts that PVT data of Exp-6 systems obey the exp-6 isotherm of eq 8; namely, $(Z-1)V^2$ versus $\rho^{-7/3}\exp\left[\alpha\left(1-C\rho^{-1/3}/r_m\right)\right]$ is linear over the whole range of densities. The procedure we followed is a direct application of exp-6 pair potential function. The result is in an equation of state while it predicts the temperature dependence of linear parameters A and B. Generally, for the establishment of the linear regularity we need not to consider details of the potential energy function of the liquid quantitatively. However, this would be the case for further quantitative treatment for molecular parameters.

The value of $\alpha$ for normal fluids is known to be between 12 and 14.[11,12] The exact value of $\alpha$ can be determined either by empirical methods or semiempirically by inverting experimentally the measured equilibrium or transport properties.[10] For cesium, we have determined $\alpha$ by inverting the second virial coefficient calculated by diatom fraction.[18] The fair accuracy of the second virial coefficient limits an absolute assignment of $\alpha$ and by a further fixation $\alpha = 9.0$ results in best exp-6 isotherms. To our knowledge, this is the lowest value assigned to $\alpha$ that surprisingly is applicable fairly well to liquid cesium.

In the exp-6 isotherm, $r_m$ is known as a parameter that slightly changes with temperature for dense fluids. For liquid metals, as the temperature is increased (or the density is decreased), the intensity of the first peak of pair correlation function $g(r)$ is diminished, its width is broadened, and its position is increased.[19,20] These are parallel to the decrease in the average coordination number and to the increase in the average nearest neighbor distance proportional to the position of the first peak. However, in the case of metals for which in the (metal-nonmetal) transition range, the nature of interaction potential does change drastically with temperature, the average number of coordination number becomes markedly smaller, but the average nearest-neighbor distance steadily increases.[20] Therefore, one has to consider compensating for the effects of change in electronic structure and its underlying interplay to the thermodynamic state of metals. We have considered these effects in the exp-6 isotherm by including the value of $r_m$ as a function of temperature. This has been made possible by using the Lennard-Jones (12-6) potential function as a pre-assumed model potential and manipulating an isotherm identical to the original LIR.[5] But, the parameters of this linear isotherm (called LJ isotherm hereafter) are in terms of the molecular parameters, the well depth $\varepsilon$ and the hard-sphere diameter $\sigma = (2)^{1/6} r_m$. Using this isotherm, we have estimated the position of minimum potential $r_m$ of the liquid cesium in the temperature range $350\,\text{K} - 2000\,\text{K}$ and pressures 50-600 bars.

At low temperatures, $r_m$ varies linearly with temperature from which the extrapolation to $301\,\text{K}$ (the melting temperature of Cs) results in $r_m (= 5.336\,\text{Å})$ that is quite close to the literature value[2] of $5.40\,\text{Å}$. (See section 4.2 for details and the graph.) Figure 2 shows the Epx6-isotherms for cesium by taking $r_m$ as a constant equal to $5.40\,\text{Å}$. The isotherms are linear up to $1800\,\text{K}$ (shown the range $1100\,\text{K} - 2000\,\text{K}$ only), and deviations are seen in the range $1800\,\text{K} - 2000\,\text{K}$. The linear range is overlapping about $400\,\text{K}$ with the metal-nonmetal transition range. Figure 3 shows the same isotherms except for $r_m = r_m(T)$ calculated by using LJ-isotherm. As it can be seen, all exp-6 isotherms satisfy the linearity limit in the range $350\,\text{K} - 2000\,\text{K}$ quite well.

LJ potential function suitably describes the interactions between the molecules of a fluid under the condition that it behaves as a normal fluid. Exp-6 potential function with $\alpha = 13-14$ numerically is almost identical to LJ potential and is also applicable to normal fluids. However, since $\alpha$ fixes the slopes of attractive and repulsive branch of exp-6 potential, as $\alpha$ decreases the exp-6 represents the potential function of a fluid with a softer repulsion and more attraction. Now, if we consider a pure fluid under the condition that the distance of nearest-neighboring atoms r is about the same as $r_m$, there is no (numerical) difference in the final results (say the isotherm) if either potential function is used to describe the interaction potential in the fluid. This is because at $r's \approx r_m$ the slopes of the two potential functions are almost the same. However, under the condition of high liquid densities ($r's < r_m$) or low liquid densities ($r's > r_m$) LJ and exp-6 potentials would result isotherms at different level of accuracies for obvious reasons (see Figure 1). The PVT data of cesium used in this study is belongs to the liquid densities corresponding to $0.989 r_m < r < 1.406 r_m$. Therefore, one would expect that at high liquid densities, the two potential functions describe the thermodynamic state of liquid cesium at the same level of accuracies, while on the contrary at low densities, the level of accuracies of results fairly depend on the how accurate the electronic interaction is described by the potential function (see next paragraph).

LJ-isotherms are shown in Figure 4 for comparison and for discussing two important points about the characteristic properties of alkali metals, especially Cs. First, the poor linearity of LJ-isotherms in the range $1350\,\text{K} - 2000\,\text{K}$ corresponding to the range in which the transition occurs is evident enough to conclude that the exp-6 potential rather than the Lennard-Jones can model the thermodynamic properties of molten cesium more accurately (see Figures 2 and 3). This conclusion is consistent with the softer short-range interaction and the larger attraction at long range of exp-6 than the Lennard-Jones. Recalling that in the expanded liquid alkali metals at high temperatures, atoms would be more polarized and thus, the interatomic interaction is conducted under the influence of electronic dipole moment of polarized atoms as well. Second, application of $r_m = r_m(T)$ improves the linearity of exp-6 isotherms over the whole range of temperature especially in the range $1800\,\text{K} - 2000\,\text{K}$ (see Figure 3). Indeed, the application $r_m = r_m(T)$ compensates for effects due to increase in the average nearest neighbor distance and actually rises the exp-6 isotherms to a self-consistency. Because $r_m$ or $\sigma$ are not explicit parameters in the LJ isotherms, their effects would appear as deviations in the linearity of isotherms, especially in the transition range where $r_m$'s changes nonlinearly with temperature. This nonlinear variation of $r_m$ with temperature is not very apparent from the pair correlation function.[16]

In Figure 5, graphically we have compared $R^2$ (linear correlation coefficient squared) values of three isotherms over the whole range of temperature. At low temperature they are about the same but at high temperatures they branch off. Deviations for LJ-isotherm cross the linearity limit (i.e., $R^2 = 0.995$) at much lower T than the other two exp-6 isotherms. The linearity completely persists up to and at $T_c (= 1924\,\text{K})$[19] of Cs when $r_m = r_m(T)$ is applied to exp-6 istherms.

Here it is of interest to note that density dependence on the structure factor of liquid cesium have been shown to be enhanced at $1373\,\text{K}$ and $1673\,\text{K}$ markedly[21], which are belong to reduced temperature $\Delta T/T_c$ 0.29 and 0.13, respectively. Nonmetal fluids like Ar do not show such a marked enhancement at such a distance from the critical temperature. Such enhancements have been a matter of considerable controversy. One explanation of such effect is that it reflects the strong density dependence of the attractive of effective inter-ionic interaction if screening is reduced as the metal-nonmetal transition is approached. Interestingly, the region of density where these enhancements occur is the same as that in which magnetic data indicate the presence of many-electron correlation effects[21]. Now, although $R^2$ is not a physical property itself, it can be used as a parameter to estimate the level of accuracies that LJ and exp-6 potential functions analytically describe the interatomic interactions in molten cesium metal over the whole liquid range. We can see that enhancements in $R^2$ occur (see Figure 5) at $1350\,\text{K}$ and $1650\,\text{K}$ markedly (and at $600\,\text{K}$ and $900\,\text{K}$ barely). Regarding the foregoing discussions, it can be concluded that the LJ potential function trust less attractions than exp-6 potential because the enhancement in $R^2$ is overwhelming for LJ-isotherms in the expanded liquid.

Since the principal contours of the $R^2$'s versus T at all T's specially at high temperatures are very similar, the interplay between electronic interaction and the thermodynamic properties are modeled by exp-6 and LJ potential functions at different levels of accuracies.

**4.2. Linear Parameters.** The linear parameters A and B of exp-6 isotherms depend upon temperature and upon the parameter $\alpha$ (see eq 5 and 9). The value of $A_2$ is so small that $A \approx -A_1/RT$. Since $A_1$ is related to the attractive of the exp-6 pair potential function, it is likely that A has the same role as second virial coefficient.[4,5] For the three isotherms of liquid Cs, values of A have plotted versus temperature, depicted in Figure 6, from which several important facts about Cs can be argued. First, the values of A for all the three isotherms are negative, and $|A|$ of the two exp-6's are larger than LJ's. This clearly indicate that exp-6 potential function trusts a larger attraction to characterize the thermodynamic behavior of the liquid Cs both at low and at high temperatures. Second, the value of $|A|$ is decreased with temperature, and shows a (broad peak) maximum at high T's. For LJ-isotherm it occurs almost at $1400\,K$ whereas for exp-6 isotherms with and without $r_m = r_m(T)$, it occurs at $1700\,K$ and $1750\,K$, respectively. The values of A for exp-6 isotherms are just level off at high T's. From the insert in Figure 6, it can be seen that, in the case of LJ potential, the turnover in the variation of A versus T is enhanced. The turnover in Figure 6 may be interpreted by considering the equilibrated liquid-vapor as the detailed in the following. In the low-temperature region there is an extensive overlap of the valence electron orbitals in accord with the nearly free electron (NFE) model to which no appreciable correlation exists between a given electron and its respective ion. Thus, modes of interaction of electron spin are unimportant. However, as the temperature is increased the diversion of the nature of forces leads to the structure forming clusters of different sizes that my be viewed as the overlap of a certain number of localized orbitals, say 2, to form a dimer. The experiment and theory state that a partially ionized dimer could be formed from penetration of one ground state atom into the valence shell of another highly excited atom of alkali vapor, gaining some stability due to orbiting the single valence electron around the dimer. It can be assumed that there are clusters in liquid state (at high temperatures) corresponding to the clusters of equilibrated vapor. [This comes to be true by the fact that diatom fraction in the vapor of alkali metals is also increased with temperature.[22]] Therefore, at high enough temperatures two potential paths of singlet and triple type interaction between the correlating valence electrons are possible. The singlet potential function is attractive with a deep well while the triplet one is of the repulsive nature with a shallow potential well. Although, the triplet type interaction is energetically less stable than the singlet, statistically it is more favored. Such effective interactions can be estimated by the second virial coefficient $B_2(T)$. Therefore, the second virial coefficient is contributed from singlet and triplet states, and by using the statistically weighted relation for the overall second virial coefficient

$$B_2(T) = \frac{1}{4}\left[^1B_2(T) + 3\,^3B_2(T)\right], \qquad (10)$$

the augmented second virial coefficient can be manipulated as

$$B_2(T) = \frac{\pi N}{2}\int_0^\infty \left[\left(1 - e^{-^1u(r)/kT}\right) + 3\left(1 - e^{-^3u(r)/kT}\right)\right]r^2 dr \qquad (11)$$

where the superscripts 1 and 3 designate singlet and triplet states, respectively, and all other symbols have their usual meanings. The statistical weighting factors $1/4$ and $3/4$ have been applied to singlet and triplet $B_2(T)$, respectively. The structure $^1u(r)$ and $^3u(r)$ are different and, therefore, the roots of singlet and triplet integrands (eq 11) occur basically at different temperatures so that $B_2(T)$ may not have roots at any temperature. Stating this in another way, the value of $B_2(T)$ at any temperature is so counter-balanced by the contribution of singlet and triplet type interactions that eq 11 may not have a root, and thus the Boyle temperature, e.g., where $B_2(T) = 0$, cannot be determined. Now, considering such properties of alkali metals, one may predict that the variation of $B_2(T)$ for vapor with temperature goes through a maximum and does not cross the temperature axis, and probably spans only in the negative region. Since the extent of singlet and triplet formation has not a sharp cut off, the shape of A at the maximum is structureless and more likely appears as a plateau. Now, since both vapor and liquid (in the transition region) are mixtures of ploy-atoms and their extents are functions of temperature, on the temperature-composition diagram, any composition in the liquid has a corresponding composition in the vapor. Therefore, as second virial coefficient is defined for a vapor, it can be proposed that the parameter $A$ represent a similar attractive interaction in the corresponding liquid. Recalling that the parameter $A$ of the LIR has already shown to behave similar to second virial coefficient of normal (nonmetallic) fluids,[5] the small turnover of $A$ in Figure 6 can thus be attributed to counter balancing effects due to the singlet and triplet type interactions, which are developing by the development of electron correlation in the high-temperature region, where the dimer (and trimers, etc.) in liquid Cs starts to form. Further investigations will be made to prove these arguments rigorously.

On the other hand, as the temperature is increased the attraction in general is decreased. But, as an alkali atom becomes appreciably polarized, the attraction somewhat is increased. Such a behavior is enhanced because a higher attraction is associated with singlet formation.

It should also be noted that, however, there is no guarantee that highly polarized attractive alkali metal atom still obey a radial force field. The effects of such changes in the force fields must also be regarded in the main discussions of the linear isotherms and thus on the behavior of the parameter $A$.

**4.3 Molecular Parameters.** Structure change can be rendered more evidently by examining the Figure 7 which depicts a plot of $r_m$ versus temperature, extracted from LJ-isotherms of liquid Cs, in the range $350\,\text{K} - 2000\,\text{K}$. Up to $1350\,\text{K}$, $r_m$ increases linearly and bends upward thereafter with small oscillations in the range $1650\,\text{K} - 1750\,\text{K}$. These observations up to $1350\,\text{K}$ are in accord with results extracted from the pair correlation function[19] which predicts a steadily increase in the average nearest neighbor distance up to $T_c$. On comparing our observation in this study (e.g., $r_m = r_m(T)$) with the result from the pair correlation function, a doubt arieses that the integration of first peak of pair correlation function for the coordination number and for $r_m$, especially at high temperatures, is exact. This can be more argued by the fact that as the temperature is increased the peaks get broader and diminish in height which is an indication of high fluctuations in density characteristic of the transition.

Figure 8 shows the plots of $\varepsilon/k$ versus temperature for liquid Cs. Notice in the model described in this study, $\varepsilon$ is the binding energy of molecules 1 and 2 in the ensemble of a similar N-2 molecule. The most effective interaction of an atom at the center is with the first shell of the body-centered lattice containing eight atoms. The molecular parameter data of the plots in Figure 8 were calculated from exp-6 isothems using relations in eq 5. With $r_m$ constant (lower plot), as the temperature is increased the well depth $\varepsilon$ is decreased steadily up to $1350\,\text{K}$ and thereafter it is increased rather sharply. These observations substantiate an interplay between the change in electronic structure, as the transition occurs, and the thermodynamic properties. It is not hard to believe that the minimum in Figure 8 is occurred due to the nonlinear change in the nearest neighboring distance with temperature (see Figure 7). The broad structureless minimum indicates the transition from metallic character is structureless. Just to show that $\varepsilon/k$ very much depends on the average nearest neighbor we constructed the upper plot in Figure 8. This plot is the same plot but $r_m = r_m(T)$ is used to perform the isotherms and then to calculate $\varepsilon/k$. Comparison of plot of the two plots clearly demonstrates, as is also well-known from theoretical and experimental neutron scattering studies,[23] that the inter-ionic interaction in metals, which determines the arrangement of ions or the ionic structure and the average nearest-neighbor as well, is the effective factor in the sense that it depends on the density.

According to calculations based on exp-6 isotherm, the binding energy of Cs in liquid state is 0.586 eV at $350\,K$. By accounting the effective interactions of (only) 8 closet neighboring atoms in a body-centered unit cell for liquid Cs, the binding energy of a pair of atoms would be 0.0733 eV. Recent ab initio calculations show that[24] the binding energy of Cs vapor is equal to 0.443 eV (for singlet state) and 0.035 eV (for triplet state). Then, the statistically weighted average of the lowest states would have a binding energy of 0.137 eV. Therefore, there is 46% excess binding energy in an isolated pair of Cs atoms compared to binding energy of a pair in compressed liquid Cs. Since the equilibrium interatomic position is not changed appreciably, a pair of Cs atoms are experienced more repulsion by the presence of the other N-2. Therefore, the parameter of pair potential of a real dense system would be different from that of an isolated one.[17] As the temperature increases, the binding energy seemingly decreases and then turns overs upward where the polarization appreciably increases, leading to a higher binding energy.

In all cases, the liquid density data at high pressures up to $1950\,K$ were taken from ref 7 and at $2000\,K$ taken from ref 8. Table 1 shows the results of the applications of exp-6 isotherm (eq 8) to the molten cesium in the range $350\,K - 2000\,K$. The pressure ranges, density ranges, and $R^2$ values are also included. The linear parameters of the LJ isotherms are shown in the Table 2. The results for exp-6 isotherms with $r_m = r_m(T)$ are shown in Table 3.

The exp-6 isotherm applies completely well to Li through Rb. Available liquid PVT data for these metals are limited to a density range well above their characteristics density limits.[8,25,26] Further data in the range where metal-nonmetal transitions have occurred are required to check the range of accuracies of application of exp-6 potenial, as we have shown in this study by examining the exp-6 isotherm, to these metals.

In general, liquid metal may be thought of as two intermixture fluids, one composed of ions and another composed of conduction electron gas. The theoretical difficulties in obtaining a self-consistent treatment would be enormous if they are treated separately. Also, since the electron gas only exists when many metal atoms are brought together to form a dense liquid, many body forces are present, and at first sight a pair potential description would seem inappropriate. Fortunately, the assumption of pseudo-atom reduces the problem of interatomic forces to one of essentially the same type as are considered for insulator fluids. By the pseudo-atom inspiration, the conduction electrons distribute themselves around each ion to form a screening charge. This point of view does not imply that the conduction electrons are localized rather a part of a general charge cloud is allocated locally to each ion. This theory is not as well founded as that used for the forces in nonconducting liquids but it has advantages of being based on a clear picture of a metal and of fitting into the conventional liquid theory based on a pair potential.[27-29]

Contributions to the interatomic forces in liquid metal are from electron-electron as well as ion-electron and ion-ion interaction. Treatment of screened ion-electron interaction leads to method of pseudopotential and the interatomic potential in the final form is detailed mathematically but pictorially looks like the potential of non-conducting fluids, except for the oscillations at long range originated from the periodic force field of the ions order. Even if the pseudopotential theory were of interest for its sort of quantum mechanical basics, its final form would not necessarily produce the simple type of isotherm produced by exp-6 potential in this study. Furthermore, pseudopotential is applicable as long as the nearly free electron model can be applied accurately, hence the metal-nonmetal transition is not covered by the general method of pseudopotential. The oscillations at long range are observed from theoretical work on the experimental neutron scattering data of liquid cesium, whose amplitude is dumped out. It dumps out gradually as the temperature is increased and vanishes in the metal-nonmetal transition. We cannot detect any effect on the isotherm 8 due to the lack of an oscillation at long range of exp-6 potential probably because, as mentioned before, the PVT data of cesium used in this study is belonging to the liquid densities corresponding to $0.989 r_m < r < 1.406 r_m$ and thus do not extend to the oscillation region range. Molecular dynamic simulation of liquid sodium has shown that truncation of the so-called friedel oscillation does not modify the structure factor of liquid Na much.[27] In short, the NFE model and the pseudopotential theory are used widely in the theory of liquid metal. To our knowledge, no other molecular theoretical studies exist to allow a conclusive direct comparison of the thermodynamics behavior of liquid alkali, especially the linear regularity (eq 8) presented in this paper around metal-nonmetal transition range.

Near the triple point at the ordinary liquid density, the effective pair potential of alkali metals can be concluded by the pseudopotential theory based on the nearly free electron model.[23] Occurrence of metal-nonmetal transition implies that the electronic structures in low-density expanded liquid cesium are fundamentally different. Liquid cesium, typically, just above its freezing point possesses a large degree of correlation in the atomic position and may be considered as a normal liquid metal. Therefore, the electrical conductivity can be explained within the framework of NFE. By employing the measured values of structure factor[30] together with model pseudopotential form given by Ashcroft,[31] the electrical conductivity of liquid cesium has been reported. It was found that at densities $\geq 6\times10^{21}$ atom/cm$^3$ (belonging up to $1300\,\mathrm{K}$) the NFE model predicts experimental electrical conductivity along the coexisting line quite good.[32] For densities $< 6\times10^{21}$ atom/cm$^3$, however, the mean free path of the electron becomes smaller than the nearest-neighbor distance of atoms and the deviations from experimental electrical conductivity become progressively large. The general failure of NFE is attributed not to the breakdown of this model, rather the NFE breakdown more likely reflects the importance of electron correlation below the density limit. Therefore, as the electron correlation significantly increases the pseudopotential, becomes invalid. In other words the pseudopotential describes and quantifies the liquid alkali metal only where metallic characters persist. The consequence of change to the nonmetallic character is the production of an overwhelmingly large number of polarizable atoms and small clusters, which derive the potential function, at long ranges, into a more attractive regime as compared to a normal metallic behavior. Accordingly, the exp-6 potential turns out to be rather valid because it presents an attractive potential more than the simple dispersive attraction characteristic of a normal fluid. The model and the theory in this study present the exp-6 isotherm regularity that is valid up to about $124\,^\circ\mathrm{C}$ below the $T_c$ and up to $T_c$ by applying a correction by using $r_m = r_m(T)$.

From the foregoing discussions, we see that at the onset of metal-nonmetal transition, concluded from the behavior of electrical conductivity as the critical temperature is approached,[3,20] the linear behavior expected from LJ isotherm turns to a quadratic behavior. Application of exp-6 potential with $\alpha = 9.0$ yields a linear isotherm formulation that persists linear over the whole liquid range. We attribute these observations to the characteristics of exp-6 potential function at long range that describe, quantify, and compensate for the progressive attractive interaction resulting from generation of polarizable atoms and clusters as the critical point is approached, as if cesium liquid metal behaves like a normal liquid over the metal and nonmetal ranges.

Therefore, in the transition region no reliable methods exist to derive an effective interaction potential function. This remains as a feature to investigate in order to consider the drawbacks of the application of exp-6 pair potential to alkali metals in the transition region, which requires a characteristic interaction potential.

## 5. Conclusions

In this study, in the light of causes of metal-nonmetal transition in liquid alkali metals, it is shown that their thermodynamic effects can be circumvented by the employment of exp-6 potential function and fully applied to liquid cesium.

At the onset of transition, delocalized electrons become relatively localized and clusters of different sizes with weak van der Waals interparticle interaction are formed.[3,6] Since the liquid Cs in the transition region is a mixture of monoatomic and ployatomic particles, the exp-6 potential function is used to establish the linear isotherms EOS that are valid up to $1800\,\mathrm{K}$ (corresponding to the density $0.842\,\mathrm{gr}/\mathrm{cm}^3$ at 90 bar). Application of $r_m = r_m(T)$ is a direct check for the effects of change in structure on the linearity of isotherms, especially in the transition region, which is exclusively feasible using exp-6 isotherm. When $r_m = r_m(T)$ is used quantitatively, the application of exp-6 isotherm to liquid Cs is extensively improved in the range $350\,\mathrm{K} - 2000\,\mathrm{K}$, especially in the transition region.

The linear exp-6 isotherm EOS indeed employs a versatile (attractive) potential and has included the parameter of the equilibrium interatomic position $r_m$. These relate, respectively, the cause of linear LJ isotherms breakdown in the transition region to the development of weak van der Waals interatomic forces as separated atoms plus clusters are formed and to the (nonlinear) increase of average nearest-neighbor distance.

Exp-6 provides a means to remove singularities in the thermodynamic properties of liquid Cs, which is observed as the results of change in electronic structure in the transition range.

Plots of $R^2$ versus T (see Figure 5) for LJ isotherm actually indicate that a (rather sharp) transition (around $1650$ K) could occur according to a certain thermodynamic order. However, since in the transition region, effects of electronic structure of monoatoms plus only fractions of ploy-atoms describe the thermodynamics of the system, a definite order cannot be expected. Recalling that $A$ is related to the attractions in the liquid system, these conclusions would be obviated when we see that the plot of $A$ versus T in Figure 6 goes through a broad structureless maximum. This might be the characteristic of a structureless transition with no thermodynamic order.

The exploration and advancement on linear isotherms of alkali metals in this study are assisted by the LJ isotherm. Lennard-Jones potential tactically describes the right order of long-range attractions in normal (nonmetallic) fluids, and thus it is taken as an implication to cover for the effects of the transition in alkali metals by an appreciable approach.

At the level of theory presented in this study, the thermodynamic effects caused by the polarization of atoms in expanded liquid alkali metals can be included by exp-6 potential function to demonstrate their thermodynamic properties like normal (nonmetallic, nonquantum) liquids.

**Acknowledgement**

M.H.G. thanks the Research Committee of the Shiraz University for supporting this project.

Table 1. Linear Parameters of Application of the Exp-6 Isotherm (Eq 8) to Molten Cesium Metal [a]

| T (K) | $R^2$ | $A \times 10^8$ $(m^3/mol)^2$ | B $(mol^{1/3}/m)$ | $\Delta\rho$ $(gr/cm^3)$ | $\Delta P$ (bar) |
|---|---|---|---|---|---|
| 350 | 0.9991 | -30.759 | 1363.38 | 1.815-1.880 | 50-600 |
| 400 | 0.9998 | -26.398 | 1178.30 | 1.787-1.854 | 50-600 |
| 450 | 0.9999 | -23.286 | 1046.88 | 1.759-1.828 | 50-600 |
| 500 | 0.9997 | -20.275 | 918.08 | 1.730-1.803 | 50-600 |
| 550 | 0.9997 | -18.266 | 833.57 | 1.702-1.764 | 50-600 |
| 600 | 0.9994 | -16.144 | 742.17 | 1.673-1.753 | 50-600 |
| 650 | 0.9998 | -14.495 | 671.07 | 1.645-1.728 | 50-600 |
| 700 | 0.9998 | -13.344 | 622.55 | 1.617-1.703 | 50-600 |
| 750 | 0.9999 | -12.443 | 585.27 | 1.589-1.670 | 50-600 |
| 800 | 0.9997 | -11.623 | 551.27 | 1.561-1.652 | 50-600 |
| 850 | 0.9997 | -10.760 | 514.18 | 1.533-1.628 | 50-600 |
| 900 | 0.9996 | -10.054 | 484.39 | 1.505-1.603 | 50-600 |
| 950 | 0.9997 | -9.439 | 458.55 | 1.476-1.578 | 50-600 |
| 1000 | 0.9999 | -9.002 | 441.37 | 1.447-1.552 | 50-600 |
| 1050 | 0.9998 | -8.418 | 416.30 | 1.416-1.527 | 50-600 |
| 1100 | 0.9998 | -7.901 | 393.91 | 1.385-1.503 | 50-600 |
| 1150 | 0.9999 | -7.447 | 374.53 | 1.352-1.478 | 50-600 |
| 1200 | 0.9997 | -7.067 | 358.58 | 1.320-1.453 | 50-600 |
| 1250 | 0.9995 | -6.690 | 342.45 | 1.287-1.429 | 50-600 |
| 1300 | 0.9990 | -6.370 | 329.11 | 1.254-1.404 | 50-600 |
| 1350 | 0.9985 | -6.090 | 317.56 | 1.220-1.380 | 50-600 |
| 1400 | 0.9981 | -5.885 | 310.13 | 1.186-1.355 | 50-600 |
| 1450 | 0.9958 | -5.698 | 303.59 | 1.151-1.330 | 50-600 |
| 1500 | 0.9976 | -5.570 | 300.13 | 1.115-1.305 | 50-600 |
| 1550 | 0.9970 | -5.465 | 298.37 | 1.078-1.280 | 50-600 |
| 1600 | 0.9963 | -5.365 | 296.83 | 1.039-1.255 | 50-600 |
| 1650 | 0.9960 | -5.261 | 295.01 | 0.994-1.230 | 50-600 |
| 1700 | 0.9974 | -5.039 | 284.84 | 0.976-1.205 | 100-600 |
| 1750 | 0.9971 | -5.022 | 288.94 | 0.926-1.179 | 100-600 |
| 1800 | 0.9962 | -5.049 | 296.57 | 0.873-1.154 | 100-600 |
| 1850 | 0.9948 | -5.126 | 308.55 | 0.811-1.127 | 100-600 |
| 1900 | 0.9910 | -5.292 | 327.94 | 0.742-1.101 | 100-600 |
| 1950 | 0.9811 | -5.662 | 366.49 | 0.655-1.072 | 100-600 |
| 2000 | 0.9915 | -4.995 | 319.29 | 0.792-1.045 | 200-600 |

[a] $R^2$ Is the Squared Linear Correlation Coefficient.

Table 2. Linear Parameters of Application of the LJ Isotherm to Molten Cesium Metal (The Upper Pressure Limit Is 600 bar)

| T (K) | $R^2$ | $A \times 10^{10}$ $(m^3/mole)^2$ | B $(m^3/mole)^4$ |
|---|---|---|---|
| 350 | 0.9982 | -1010.12 | 5.17E-16 |
| 400 | 0.9993 | -890.99 | 4.66E-16 |
| 450 | 0.9997 | -807.65 | 4.32E-16 |
| 500 | 0.9997 | -722.83 | 3.95E-16 |
| 550 | 0.9994 | -675.12 | 3.78E-16 |
| 600 | 0.9985 | -607.37 | 3.47E-16 |
| 650 | 0.9990 | -560.61 | 3.27E-16 |
| 700 | 0.9993 | -530.35 | 3.16E-16 |
| 750 | 0.9993 | -510.46 | 3.12E-16 |
| 800 | 0.9988 | -487.02 | 3.04E-16 |
| 850 | 0.9988 | -463.23 | 2.96E-16 |
| 900 | 0.9982 | -444.52 | 2.90E-16 |
| 950 | 0.9987 | -429.11 | 2.87E-16 |
| 1000 | 0.9991 | -420.46 | 2.88E-16 |
| 1050 | 0.9988 | -404.77 | 2.84E-16 |
| 1100 | 0.9986 | -390.85 | 2.81E-16 |
| 1150 | 0.9985 | -379.53 | 2.79E-16 |
| 1200 | 0.9979 | -370.59 | 2.80E-16 |
| 1250 | 0.9972 | -361.44 | 2.79E-16 |
| 1300 | 0.9959 | -354.42 | 2.81E-16 |
| 1350 | 0.9948 | -349.06 | 2.84E-16 |
| 1400 | 0.9937 | -347.19 | 2.91E-16 |
| 1450 | 0.9908 | -346.15 | 2.98E-16 |
| 1500 | 0.9921 | -348.07 | 3.09E-16 |
| 1550 | 0.9904 | -351.34 | 3.23E-16 |
| 1600 | 0.9885 | -355.15 | 3.37E-16 |
| 1650 | 0.9871 | -359.91 | 3.54E-16 |
| 1700 | 0.9899 | -351.55 | 3.53E-16 |
| 1750 | 0.9878 | -361.85 | 3.79E-16 |
| 1800 | 0.9843 | -375.82 | 4.13E-16 |
| 1850 | 0.9793 | -395.75 | 4.59E-16 |
| 1900 | 0.9694 | -423.67 | 5.21E-16 |
| 1950 | 0.9486 | -472.21 | 6.28E-16 |
| 2000 | 0.9757 | -400.87 | 5.19E-16 |

Table 3. Linear Parameters of Application of the Exp-6 Isotherm of Molten Cesium Metal, with $r_m$ Used as a Function of Temperature. The Upper Pressure Limit is 600 bar.

| T (K) | R2 | Ax10$^8$ (m$^3$/mole)$^2$ | B (mol$^{1/3}$/m) |
|---|---|---|---|
| 350 | 0.9991 | -29.85 | 1404.38 |
| 400 | 0.9998 | -26.04 | 1194.71 |
| 450 | 0.9999 | -23.35 | 1043.95 |
| 500 | 0.9997 | -20.65 | 900.16 |
| 550 | 0.9997 | -18.91 | 801.58 |
| 600 | 0.9994 | -16.96 | 701.38 |
| 650 | 0.9998 | -15.43 | 622.90 |
| 700 | 0.9998 | -14.41 | 566.59 |
| 750 | 0.9999 | -13.63 | 521.95 |
| 800 | 0.9998 | -12.91 | 480.52 |
| 850 | 0.9998 | -12.10 | 438.99 |
| 900 | 0.9996 | -11.45 | 404.30 |
| 950 | 0.9998 | -10.88 | 373.81 |
| 1000 | 0.9999 | -10.52 | 350.49 |
| 1050 | 0.9999 | -9.95 | 322.13 |
| 1100 | 0.9999 | -9.43 | 297.20 |
| 1150 | 0.9999 | -8.98 | 274.76 |
| 1200 | 0.9998 | -8.61 | 255.56 |
| 1250 | 0.9997 | -8.23 | 236.89 |
| 1300 | 0.9993 | -7.90 | 220.52 |
| 1350 | 0.9990 | -7.62 | 205.85 |
| 1400 | 0.9987 | -7.43 | 193.82 |
| 1450 | 0.9966 | -7.26 | 182.51 |
| 1500 | 0.9986 | -7.15 | 173.33 |
| 1550 | 0.9982 | -7.08 | 164.72 |
| 1600 | 0.9979 | -7.01 | 156.27 |
| 1650 | 0.9978 | -6.91 | 147.40 |
| 1700 | 0.9989 | -6.65 | 137.65 |
| 1750 | 0.9990 | -6.67 | 130.95 |
| 1800 | 0.9989 | -6.75 | 125.01 |
| 1850 | 0.9986 | -6.89 | 119.24 |
| 1900 | 0.9974 | -7.14 | 114.77 |
| 1950 | 0.9937 | -7.66 | 111.98 |
| 2000 | 0.9968 | -6.93 | 105.25 |

**Figure captions**

Figure 1. Comparison of exp-6 potential for $\alpha = 9, 12, \text{and } 14$ with Lenanard-Jones. Typical molecular parameters used for demonstration.

Figure 2. Exp-6 isotherms of molten cesium from 350 K to 2000 K (shown 1100-2000 K only). Thin solid lines are linear fits for the isotherms with $R^2 \geq 0.995$. The solid thick lines are trend lines for the isotherms with $R^2 < 0.995$.

Figure 3. Same as Figure 2 except for the value of $r_m$ as a function of temperature, calculated by using the LJ isotherms.

Figure 4. Same as Figure 2 except for LJ isotherms for liquid Cs.

Figure 5. Plots of $R^2$ versus temperature for the three isotherms of Figures 2-4.

Figure 6. Plot of parameter A versus T for Cs. Insert is LJ isotherm. See the text for details.

Figure 7. The plot of $r_m$ versus temperature (extracted from LJ isotherms of Cs).

Figure 8. Plots of $\varepsilon/k$ as a function of temperature for liquid cesium calculated by using exp-6 isotherms.

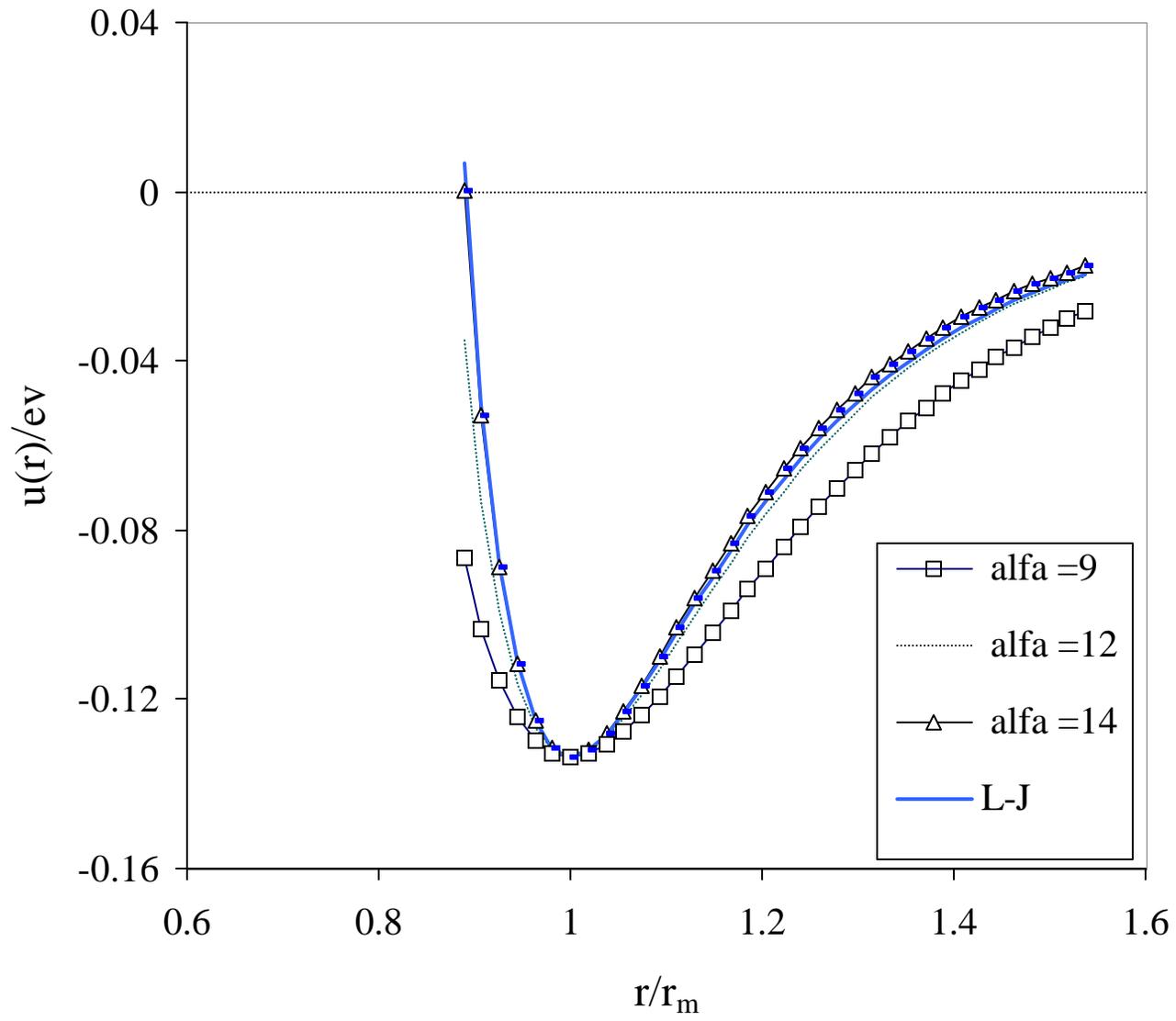

Figure 1

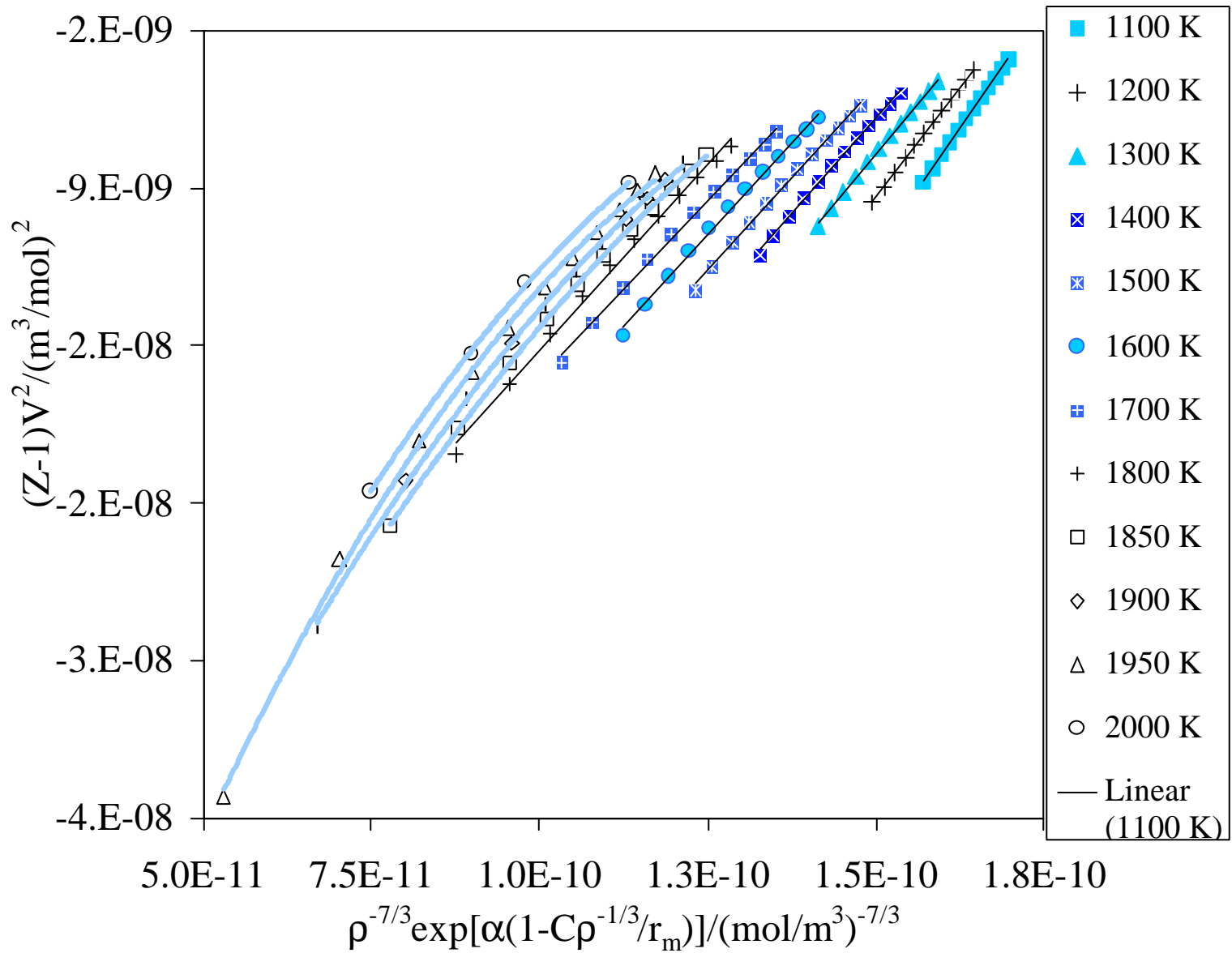

Figure 2

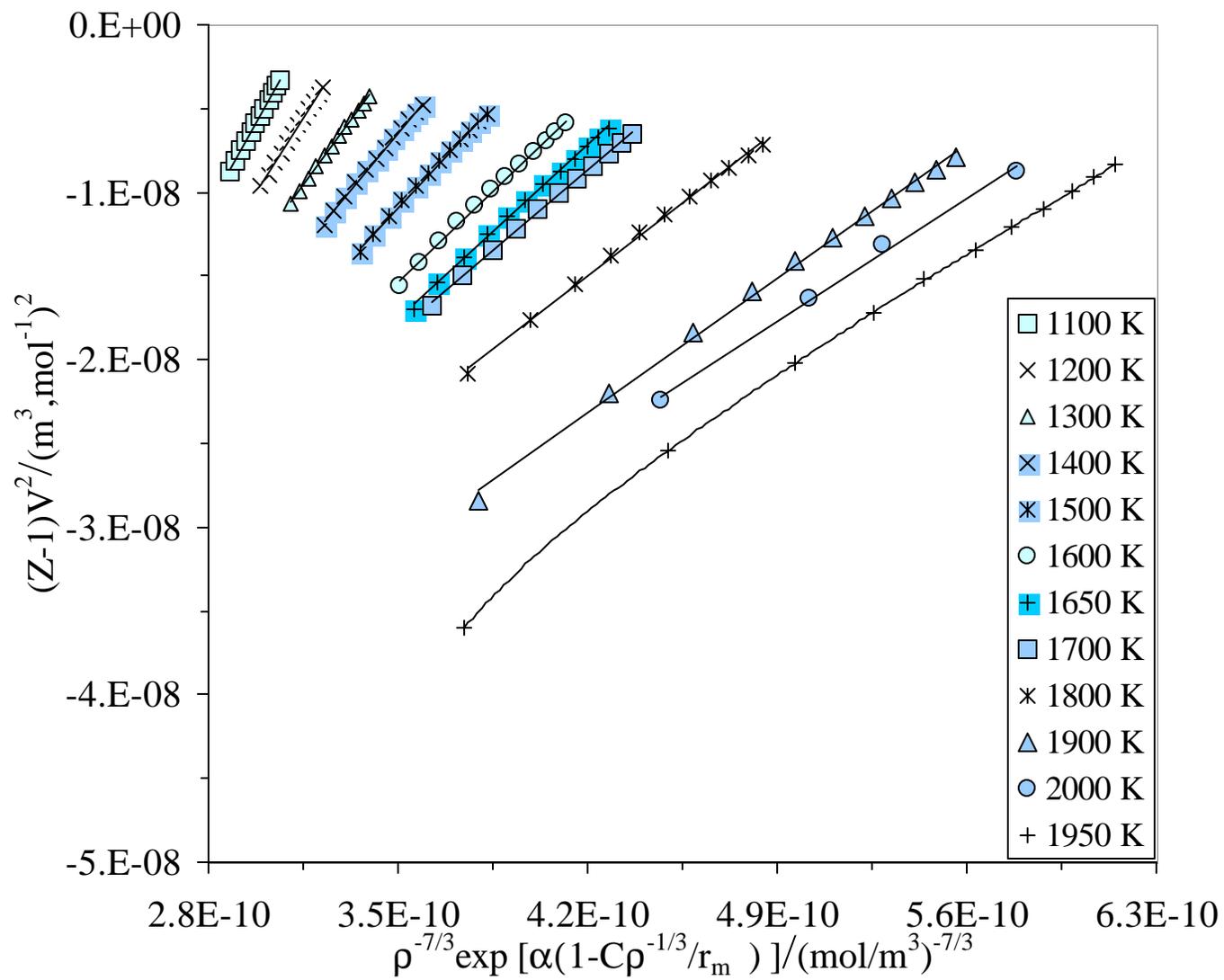

Figure 3

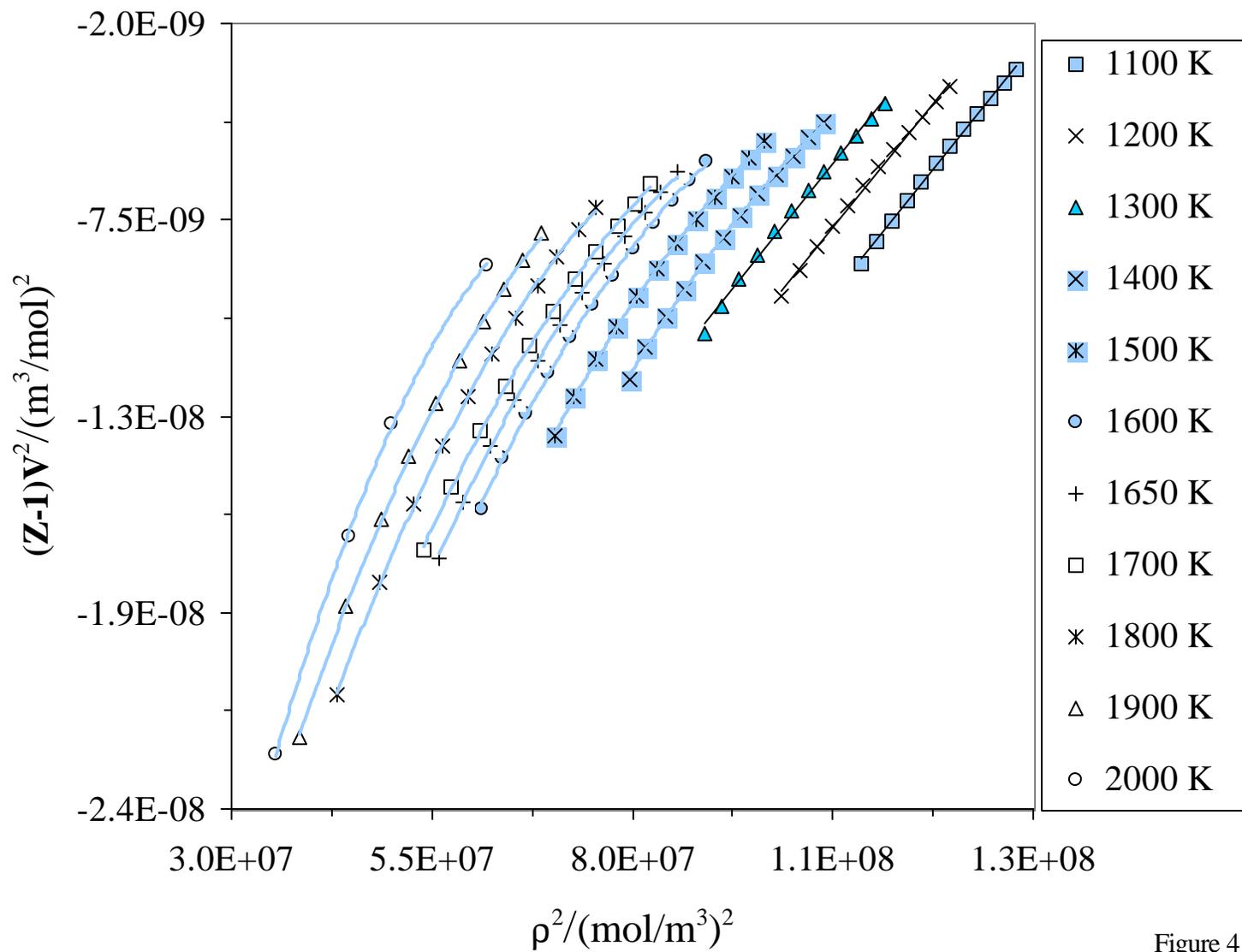

Figure 4

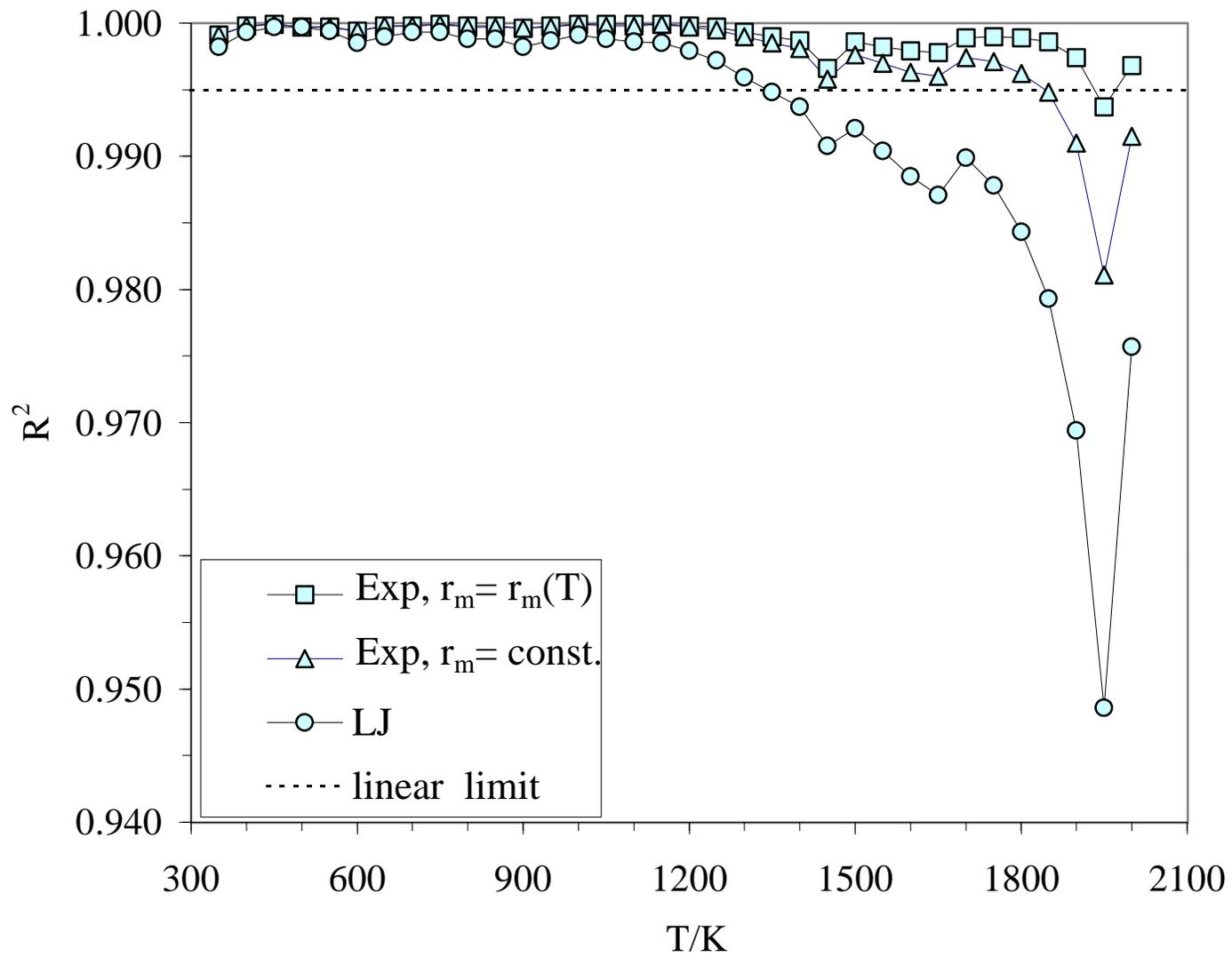

Figure 5

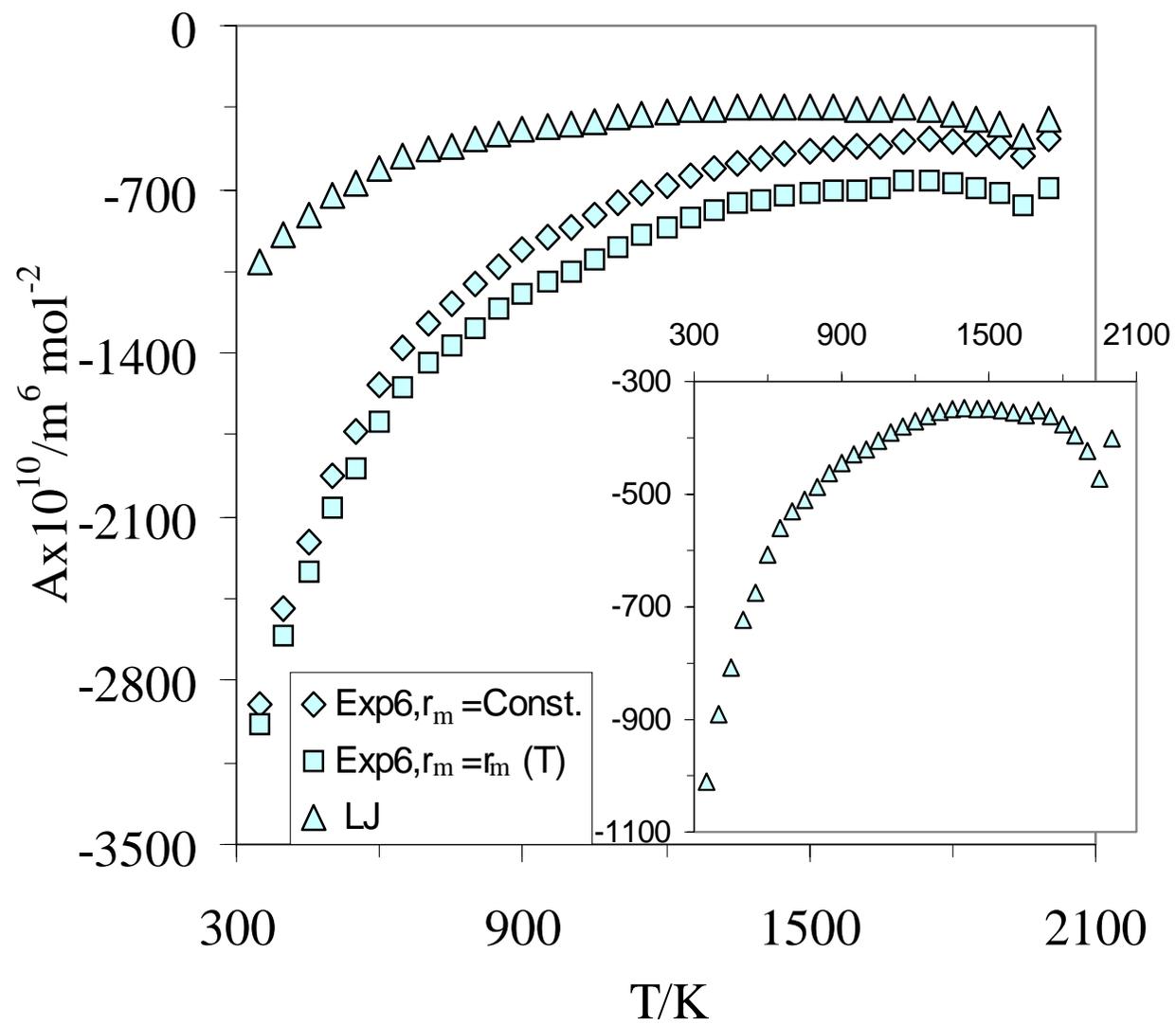

Figure 6

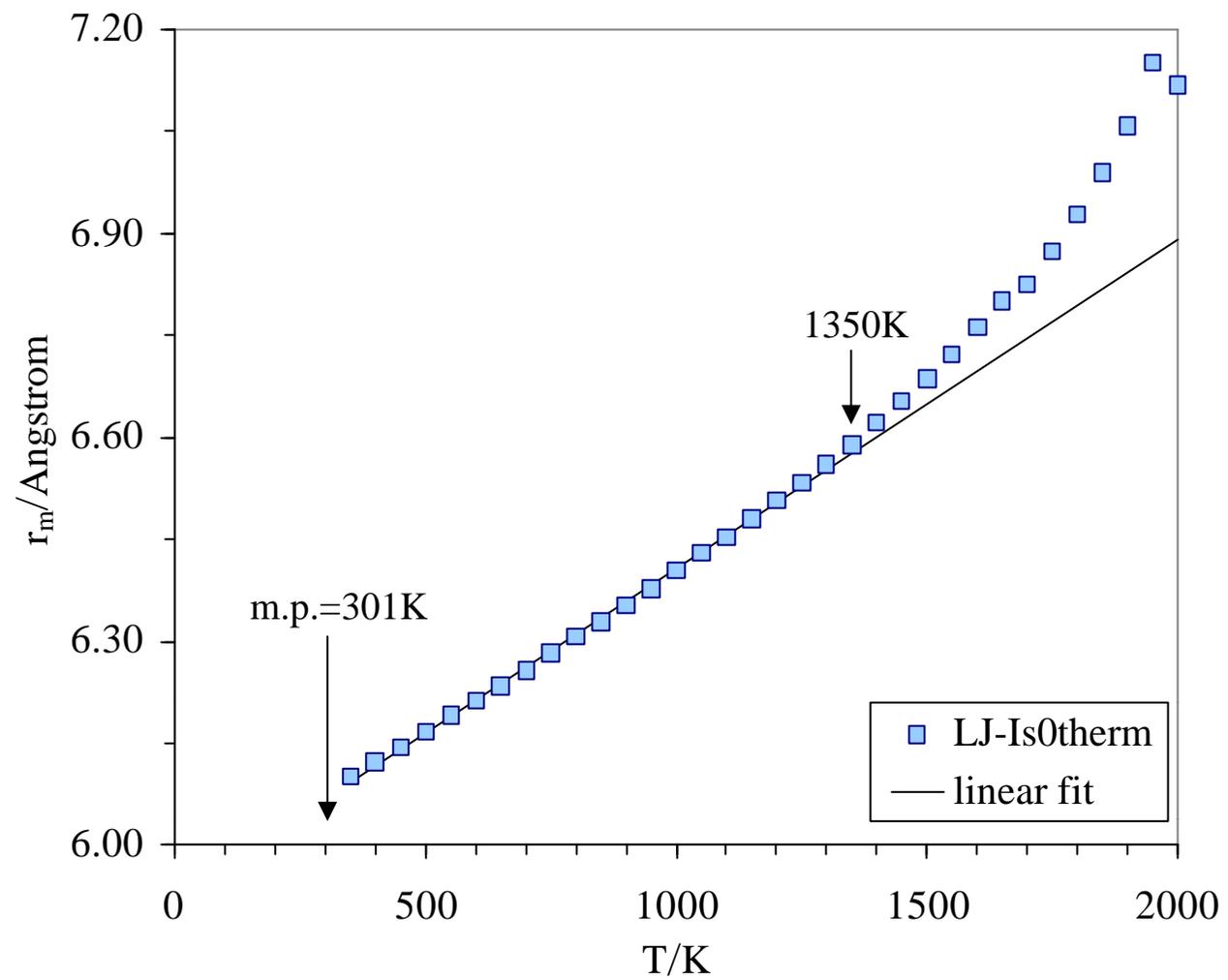

Figure 7

37

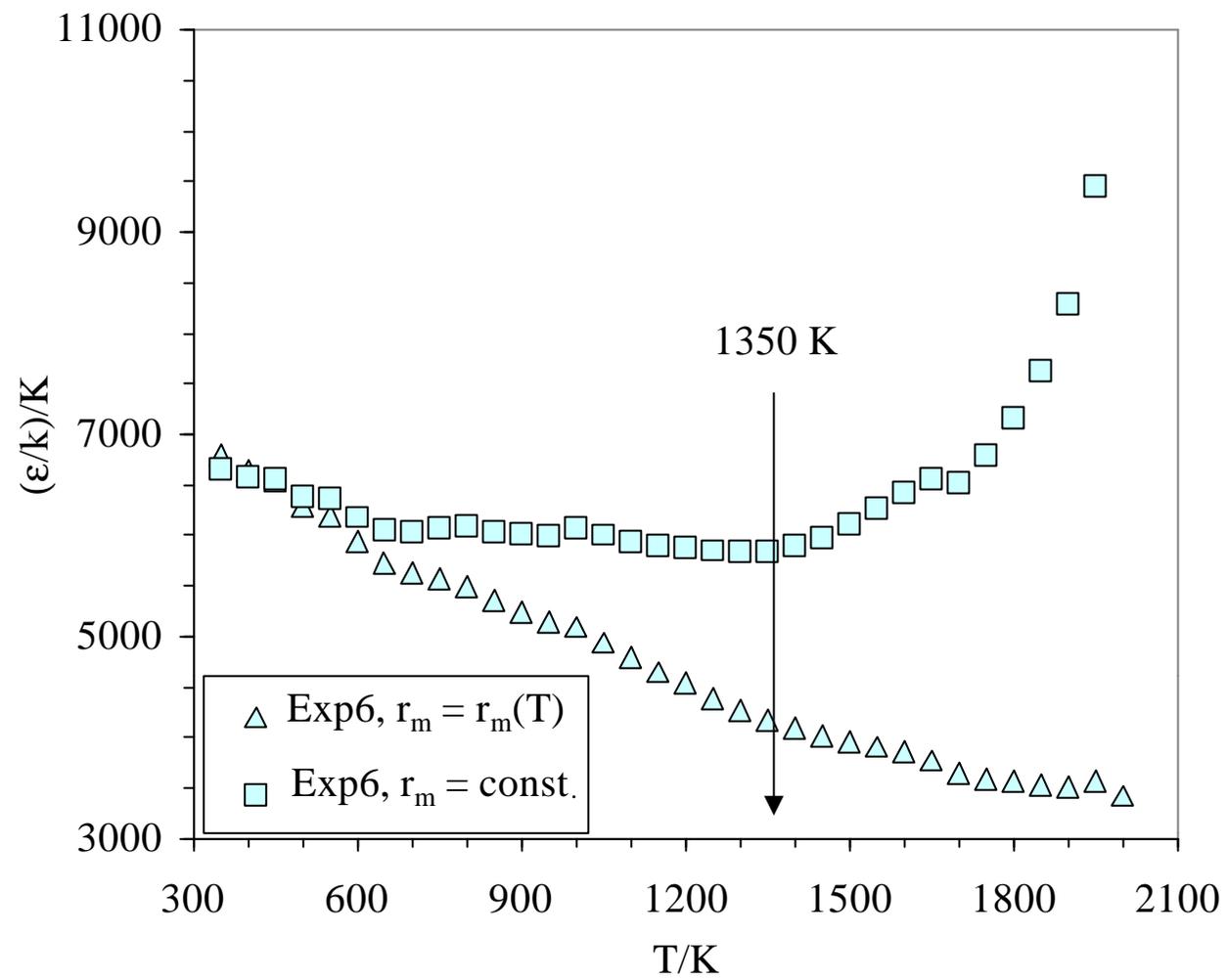

Figure 8